\documentstyle[multicol,aps,psfig]{revtex}
\begin{document}

\draft

\title{Slowly driven sandpile formation with granular mixtures}

\author{D.A.Head\cite{em1} and G.J.Rodgers\cite{em2}}

\address{Department of Physics, Brunel University,
Uxbridge, Middlesex, UB8 3PH, United Kingdom}

\date{\today}

\maketitle

\begin{abstract}
We introduce a one-dimensional
sandpile model with $N$ different particle types and
an infinitesimal driving rate.
The parameters for the model are the
$N^{2}$ critical slopes for one type of particle on top of another.
The model is trivial when $N=1$, but
for $N=2$ we observe four broad classes of sandpile
structure in different regions of the parameter space.
We describe and explain the behaviour of each of these classes,
giving quantitative analysis wherever possible.
The behaviour of sandpiles with $N>2$ essentially consists
of combinations of these four classes.
We investigate the model's robustness and highlight the key areas that
any experiment designed to reproduce these results should focus on.
\end{abstract}


\maketitle
\begin{multicols}{2}
\narrowtext

\section{Introduction}
\label{sec:intro}

Granular materials display a variety of unusual behaviour
not normally associated with
either solids or liquids~\cite{granular}.
One such example is the segregation that occurs when a
mixture of different sized granules is repeatedly shaken,
in which the larger particles
rise to the top~\cite{brazilnut,bn2,bn3}.
Similarly, a granular mixture placed inside a rotating cylinder
segregates into alternate bands along the cylinder's
axis~\cite{rotating,rotating2}.
Segregation in the absence of external perturbations has recently been
demonstrated for a mixture poured between
two vertical plates separated by a narrow gap~\cite{strat1}.
A {\em sandpile} forms in which
particles of different sizes tend to remain near the top or
bottom.
Moreover, certain mixtures also {\em self-stratify} into alternating 
layers parallel to the surface of the pile.
It is this segregation in sandpiles
that this paper seeks to address.

A sandpile is formed by the addition of particles
which then move over the surface of the pile until
finding a resting place.
Modelling this process is a highly non-trivial problem
even without the added complication of mixtures of particles.
There are two basic approaches to modelling the surface
transport of pure granular media.
Firstly, Bak~{\em et.al.}~\cite{BTW} introduced a cellular automata model
as a paradigm of their more general concept
of self-organised criticality. In this model, sequentially added particles 
can initiate a series of locally-defined topples known collectively as an
{\em avalanche}. The system is said to be {\em slowly driven} as the
timescale for particle injection is infinitely slower
than that of the subsequent avalanche~\cite{Grinstein}.
However, experimental evidence~\cite{exp1,exp2,exp3} disagrees with the
model's predicted power-law distribution of avalanche sizes, except
possibly in the limit of overdamped particle motion~\cite{ricepile}.
A second approach treats the sandpile as a continuum with
a fluid {\em rolling layer} interacting with the static 
bulk of the pile.
Analysis of the field equations of this model appears to give a greater
correspondence with the experiments~\cite{BCEP}.

Variations of both discrete and continuous
approaches have been used to try and explain the
behaviour observed in the sandpile formed by pouring a mixture
between two plates.
A discrete model, in which the particles are added in groups
and also move downslope in groups,
exhibits self-stratification
but gives no insights into the mechanism responsible for the
formation of the layers.
However, extending the continuum rate equations to incorporate mixtures
demonstrates the existence of a {\em kink}, which back-propagates
up the slope forming two layers at once, giving a possible
explanation of the experimental results~\cite{strat2}.

In this paper, and in contrast to the experiments~\cite{strat1} and their
subsequent analysis~\cite{strat2},
we consider a slowly driven system in
which particles are added individually rather than being
poured.
The entire pile is stable between particle additions and
there is no formation of a rolling layer.
The model we have chosen to study is a one-dimensional cellular
automata with mixed particle types, based on the model
in~\cite{BTW}, which has no avalanching in one-dimension.
For binary mixtures we observe four broad classes of sandpile 
structure as the relative strengths of the interactions
between the particles are varied.
Two of these classes correspond to the self-segregation and
self-stratification observed in the rolling layer case.

The algorithm for the model is described in Sec.\ref{sec:model}.
The four classes of behaviour are described and their
evolution explained in Sec.\ref{sec:results}, and the
results for mixtures with more than two particle types are also
given. In Sec.\ref{sec:disc} we discuss how experiments to
observe these classes might proceed.

\section{The Model}
\label{sec:model}

\begin{figure}
\centerline{\psfig{file=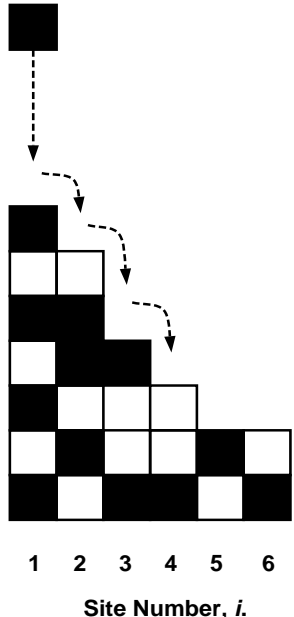,width=1.5in}}
\caption{Example of sandpile evolution for $N=2$, $z_{11}=1$,
$z_{12}=2$, $z_{21}=3$ and $z_{22}=4$. Particles of type~1 are black, and 
those of type~2 are white. For the sandpile shown above, a particle of 
type~1 added to site~1 will eventually come to rest on site~4.}
\label{f:example}
\end{figure}

The sandpile profile is described by the set of heights
\mbox{$h_{i}$, $i\geq 1$}, where all the $h_{i}$ are initially
set to zero.
An infinite wall at \mbox{$i=0$} serves as a lower bound for $i$,
whereas there is no upper bound on the values of $i$.
At each time step, a particle
is chosen from one of $N$ possible types, where each particle 
has the same dimensions and is chosen with equal probability.
A particle of type~\mbox{$\alpha\in[1,N]$} is then added to the top of
site~1 and subsequently slides to the first site~\mbox{$i\geq1$}
that obeys \mbox{$z_{i}<z_{\alpha\beta}$}, where
\mbox{$z_{i}=h_{i}-h_{i+1}$} and $\beta$ is
the type of particle currently on the surface at site~$i$.
The particle is added to the top of site~$i$,
\mbox{$h_{i}\rightarrow h_{i}+1$}
and another particle can now be added to the system.
An example of this process is given in Fig.\ref{f:example}.

The $N^{2}$ parameters $z_{\alpha\beta}$ correspond to the maximum
slope on which a particle of type~$\alpha$ can remain on top of a
particle of type~$\beta$ without sliding off.
A pure pile of just type~$\alpha$ particles has uniform slope
$z_{\alpha\alpha}$, so \mbox{$\tan^{-1}(z_{\alpha\alpha})$} can be
identified as the {\em angle of repose}. The
$z_{\alpha\beta}$ for \mbox{$\alpha\neq\beta$} are microscopically
defined quantities with no obvious macroscopic counterparts.
Generally \mbox{$z_{\alpha\beta}\neq z_{\beta\alpha}$}, since the
critical slope will depend upon which type of particle is
moving on top of the pile,
for instance types $\alpha$ and $\beta$
may have different densities.
Due to the complex nature of granular surface-to-surface
interactions it is unclear how much of this parameter space
corresponds to physical reality.
The labels for each particle type are just dummy variables, so
without loss of generality we choose to fix
\mbox{$z_{11}\leq z_{22} \leq\ldots\leq z_{NN}$}.

\section{results}
\label{sec:results}

For $N=1$ the system reduces to the original model which
is trivial in one dimension~\cite{BTW}. For $N=2$ and $z_{11}<z_{22}$
there are four classes of solution, which
we label I to IV.
Each class can be identified
according to the {\em domain stability} of each particle type,
which is defined as follows.
A compact region of sites with
surface composition of particle type~$\alpha$ is stable
if it can reach a uniform slope $z_{\alpha\alpha}$ such that particles of 
type~$\beta\neq\alpha$ will slide through the region -- that is, if
$z_{\beta\alpha}<z_{\alpha\alpha}$.
Unstable regions can only form in instances where there is no
incoming flux of particles of the other type,
which may occur towards the right-hand side of the sandpile.
We now describe each of the four classes in turn.

\begin{figure}
\centerline{\psfig{file=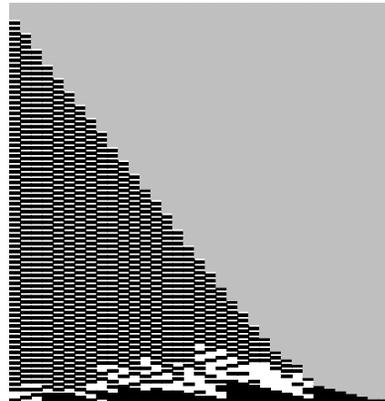,width=2 in}}
\caption{Example of a class I sandpile with the parameters $z_{11}=1$,
$z_{12}=5$, $z_{21}=7$ and $z_{22}=3$, which exhibits
{\em periodic mixing}. Particles of 
type~1 are shown as black, particles of type~2 are shown as white.
The pile given here is small, just 35 sites wide at its base, to
show the mixing clearly.
In this and all subsequent cases, simulations have been extended
up to $O(10^{5})$ particles without any observed
deviation from the characteristic behaviour. 
Note that we have rescaled the $y$-axis
to give a roughly square picture.
}
\label{f:ex_pm}
\end{figure}

{\em Class I : $z_{21}>z_{11}$ and $z_{12}>z_{22}$.}
Neither particle type can form stable domains and particles will 
usually come
to rest on particles of the other type.
We call this {\em periodic
mixing}. An example of such a sandpile is given in Fig.\ref{f:ex_pm}.
This behaviour can be demonstrated by the following single-site 
analysis. A site with an uppermost particle
of type~$\alpha$ and slope $z_{i}$ is represented by
$(\alpha,z_{i})$. Then the transition amplitude for a particle of
type~$\beta$ coming to rest on this site,
\mbox{$(\alpha,z_{i})\rightarrow(\beta,z_{i}+1)$}, is given by
the particle addition operator $P_{\alpha\beta}(z_{i})$
defined by

\begin{equation}
P_{\alpha\beta}(z_{i}) = \theta(z_{\alpha\beta}-z_{i}),
\end{equation}

\begin{figure}
\centerline{\psfig{file=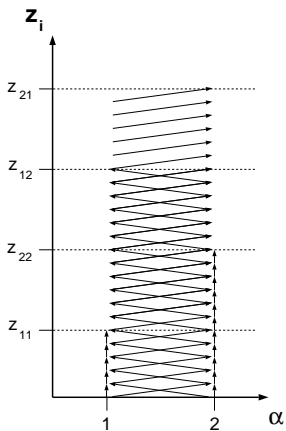,width=1.5in}}
\caption{Phase portrait of the particle addition operator
\mbox{$P_{\alpha\beta}(z_{i})$}
in the case $z_{11}<z_{22}<z_{12}<z_{21}$, where
each arrow corresponds to
\mbox{$P_{\alpha\beta}(z_{i})=1$}, that is, an allowed particle addition.
$\alpha$ denotes the type of particle on top and
$z_{i}$ is the slope.}
\label{f:phase}
\end{figure}

\noindent{where
\mbox{$\theta(x)=1$ for $x>0$ and $\theta(x)=0$ for $x\leq0$}.
Low values for $z_{i}$ are transitory, but for large
$z_{i}$ added particles slide through, possibly coming to rest
on site \mbox{$i+1$} and reducing $z_{i}$ by 1.
Hence the bulk properties of the sandpile will be characterised
by the action of $P$ in the region of large $z_{i}$.
An example is given in Fig.\ref{f:phase} for the case
\mbox{$z_{11}<z_{22}<z_{12}<z_{21}$}, which clearly shows
periodicity for $z_{i}$ near
the maximum slope, which in this case is $\approx z_{12}$.
Note that for \mbox{$z_{22}>z_{21}$} we do not have strict periodicity
as type~2 particles are occasionally added
consecutively.
}

\begin{figure}
\centerline{\psfig{file=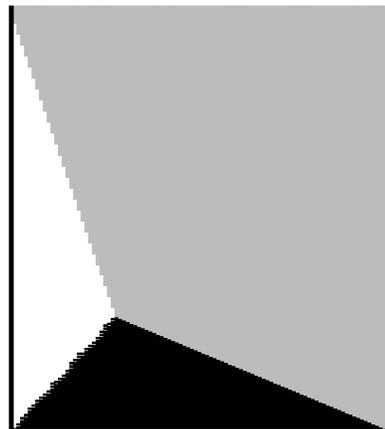,width=2in}}
\caption{Example of a class II sandpile with $z_{11}=1$, $z_{12}=3$, 
$z_{21}=5$ and $z_{22}=7$, demonstrating {\em discrete self-segregation}.
Particles of type~1 are black, those of type~2 are white.
The base of the pile is 100 sites wide.
}
\label{f:ex_dss}
\end{figure}

{\em Class II : $z_{21}>z_{11}$ and $z_{12}<z_{22}$.}
A stable domain of
type~2 particles with uniform slope $z_{22}$
builds up to the left of the 
sandpile, and type~1 particles slide through to the right
forming a domain with uniform slope $z_{11}$. The result
is {\em discrete self-segregation}, as the example in Fig.\ref{f:ex_dss}
demonstrates.
The boundary between the two domains moves to the right
as particles of type~2 come to rest on sites with slope $z_{11}$,
so if $z_{12}>z_{11}$ some periodic mixing may occur.
If $z_{21}>z_{12}$, the mixing remains confined to a narrow
layer at the boundary, but if $z_{21}<z_{12}$ it expands to the bulk
of the pile, giving a sandpile similar in appearance to case of
the periodic mixing described previously.
The self-segregation and boundary behaviour can also be seen from
the phase portraits of the particle addition operator
$P_{\alpha\beta}(z_{i})$.
We justify referring to the case
\mbox{$z_{11}<z_{21}<z_{12}<z_{22}$} as discrete self-segregation
by the observation that
the region of periodic mixing collapses to a much narrower
layer under most small alterations in the dynamical rules.
We will return to this point later when we discuss the robustness
of the model.

It is possible to construct a global
solution of this class of sandpiles.
Define $L_{B}$ and $L$ to be position of the boundary and the
right-hand edge of the sandpile respectively,
so that $h_{L}>0$, $h_{L+1}=0$, all
sites~\mbox{$1\leq i\leq L_{B}$} have type~2 particle on top and all
sites~\mbox{$L_{B}<i\leq L$} have type~1 particles on top.
A type~2 particle added to $i=L_{B}$ will
reduce the slope of site~$L_{B}-1$ by 1, so the next
type~2 particle added
will stop at $L_{B}-1$, then $L_{B}-2$, $L_{B}-3$, and
so on. Similarly, type~1 particles will be added to
$L$, $L-1$, $L-2,\ldots,L_{B}+1$, in that order.
$L$ will move to the right
by one step when $h_{L}$ increases from 0 to $z_{11}$, so in
the continuum limit

\begin{equation}
\frac{dL}{dt}=\frac{1}{2z_{11}}
\left( \frac{1}{L-L_{B}}\right) ,
\label{eq:dl}
\end{equation}

\noindent{where
the time scale has been normalised to one particle addition of either
type per unit time. The slope at $L_{B}$ increases with each type~2
particle that stops on $L_{B}$ and decreases with each type~1 particle
that stops on $L_{B}+1$, and since $L_{B}$ increases by one whenever
the slope increases from $z_{11}$ to $z_{22}$ we get
}

\begin{equation}
\frac{dL_{B}}{dt}=\frac{1}{2(z_{22}-z_{11})}
\left( \frac{1}{L_{B}}-\frac{1}{L-L_{B}}\right) .
\label{eq:dlb}
\end{equation}

\noindent{A
steadily evolving sandpile corresponds to
a constant ratio \mbox{$L_{B}/L$}, so}

\begin{equation}
\frac{d}{dt}\left(\frac{L_{B}}{L}\right)=0 .
\end{equation}

\noindent{Using this
together with~(\ref{eq:dl}) and~(\ref{eq:dlb}), we find
}

\begin{equation}
\frac{L_{B}}{L} = \frac{\sqrt{z_{22}/z_{11}}-1}
{z_{22}/z_{11}-1},
\end{equation}

\noindent{giving the average slope of the entire pile as}

\begin{equation}
\left( 1-\frac{L_{B}}{L} \right)z_{11}+\frac{L_{B}}{L}z_{22}
= \sqrt{z_{11}z_{22}},
\end{equation}

\noindent{in agreement with simulations~\cite{footnote}.}

\begin{figure}
\centerline{\psfig{file=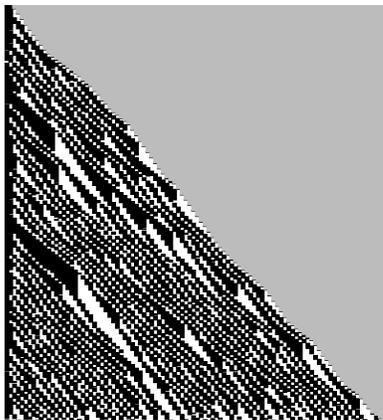,width=2in}}
\caption{Example of a class III sandpile with $z_{11}=5$, $z_{12}=15$, 
$z_{21}=1$ and $z_{22}=10$, demonstrating {\em self-stratification}.
Type~1 particles are black and type~2 particles are white.
The base of the pile is 100 sites wide.}
\label{f:ex_l}
\end{figure}

{\em Class III : $z_{21}<z_{11}$ and $z_{12}>z_{22}$.}
It might be expected that self-segregation
will also occur here, this time with type~1 particles to the left
of the boundary.
However, since $z_{11}<z_{22}$ the type~2 particles to the right
of the boundary form steeper slopes than the type~1 particles
to the left, and so now the
boundary moves upslope rather than downslope,
creating a double layer of 2's on 
top of 1's. Once the boundary reaches the left-hand wall, a thin layer of
1's quickly cover the surface of the whole sandpile,
the boundary returns to the bottom and starts propagating
upwards once more.
This moving interface corresponds to the kink described
in the rolling layer case~\cite{strat1,strat2},
and, indeed, the region of parameter space in
which it occurs is the same.
However, there are 3 significant differences between the
nature of the
self-stratification formed by this slowly-driven process and that formed
by rolling layers.
(i)~The slope of the layers
varies between $z_{11}$ and $z_{22}$,
as opposed to the uniform slope of
$z_{22}$ observed in the rolling layer case.
(ii)~The layers are narrower, typically just one particle wide.
(iii)~The rate at which the interface moves is no longer a constant but
varies according to statistical fluctuations in the
types of incoming particles.
It is even possible for the interface to stop moving altogether,
resulting in a vertical build-up of particles that is
reminiscent of a miniature self-segregated sandpile.
However, this state is unstable and the interface will
eventually start moving again, either upslope to continue
the layering process or quickly downslope to the bottom of
the pile.
Thus, layers can now start and stop in the bulk of the pile.
An example of a self-stratified sandpile is given in Fig.\ref{f:ex_l}.

\begin{figure}
\centerline{\psfig{file=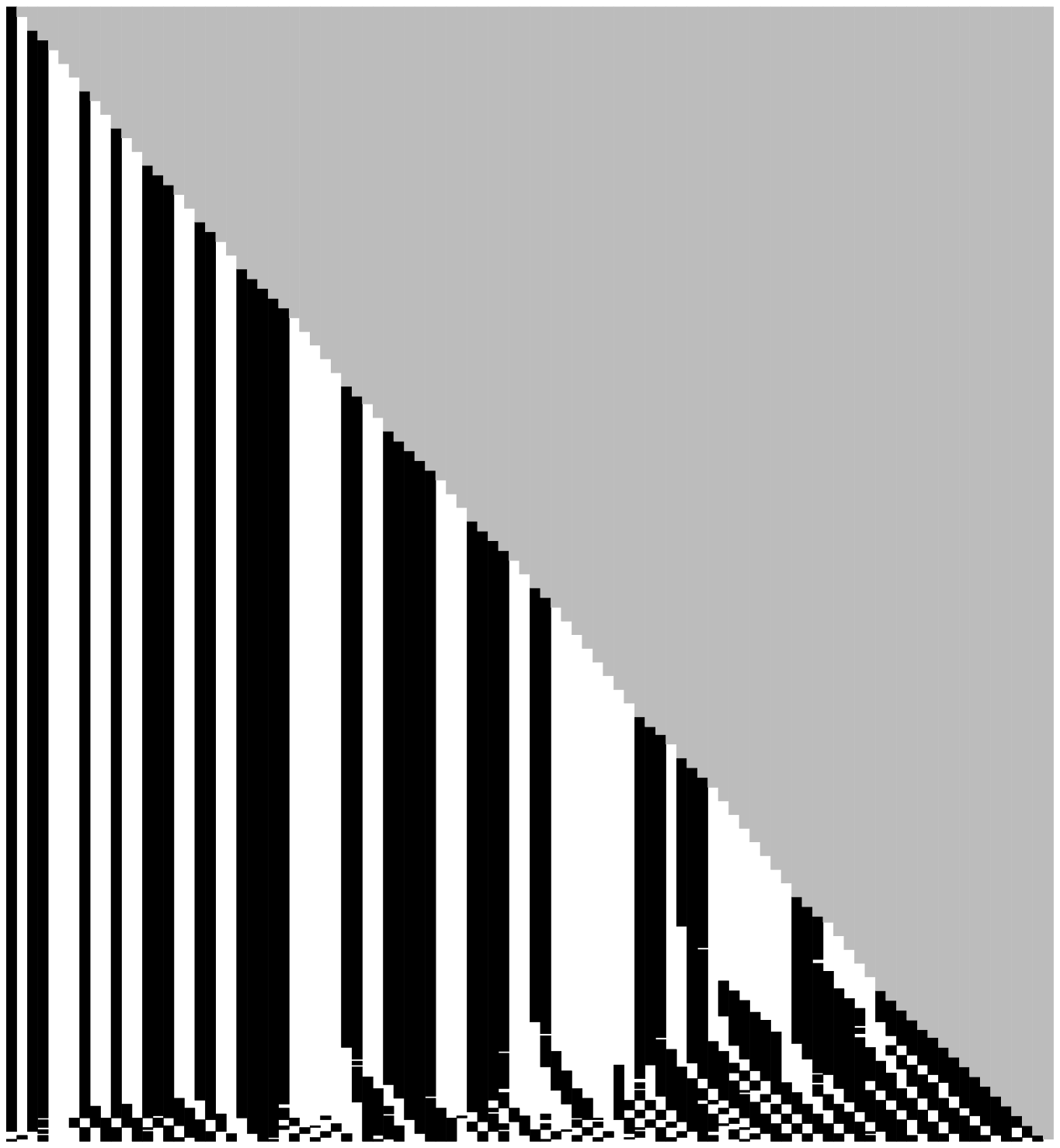,width=2in}}
\caption{Example of a class IV sandpile with $z_{11}=5$, $z_{12}=3$,
$z_{21}=1$ and $z_{22}=7$, demonstrating
{\em vertical stratification}. Type~1 particles are white, type~2 particles
are white and the base of the pile is 100 sites wide. }
\label{f:ex_vs}
\end{figure}

{\em Class IV : $z_{21}<z_{11}$ and $z_{12}<z_{22}$.}
In this final class, alternating stable
domains of 1's and 2's form parallel vertical bands.
An example of this {\em vertical stratification} is given
in Fig.\ref{f:ex_vs}.
The phase portrait of $P_{\alpha\beta}(z_{i})$
for this class demonstrates the separation
of the particle types, but further analysis requires some knowledge
of the global solution.
Once the slope of the site at the right-hand edge of any domain
decreases by one, a single layer of particles of similar type
back-propagates to the left-hand edge, when the adjacent
domain will undergo a similar process.
Thus the bulk of the sandpile
builds up a layer at a time in this piecewise fashion, from the bottom
of the pile to the top.
The process that initially generates each domain depends upon
the interaction between the bulk of the sandpile and the
qualitatively different {\em end region}
at the far right-hand side.
In the end region, a layer of 2's of thickness
\mbox{$\approx z_{11}-z_{21}+1$}
back-propagates to the first domain of type~2 particles.
The domain then broadens to the right, and the process is repeated.
Alternatively, if \mbox{$z_{12}>z_{21}$} the back-propagating
layer of 2's can be stopped prematurely by an
incoming flux of 1's, resulting
in the formation of a new stable domain of type~1 particles.
Since the particles that flow into the end region are precisely
those that did {\em not} stop in the bulk of the pile, the
number of domains of each different type tends to remain equal,
although the domains themselves get broader as the end region expands.
If \mbox{$z_{12}<z_{21}$}
the formation of new type~1 domains can still occur
when the sandpile is small
is small due to statistical fluctuations, but this
rarely occurs for larger piles which take on
a self-segregated appearance.

\begin{figure}
\centerline{\psfig{file=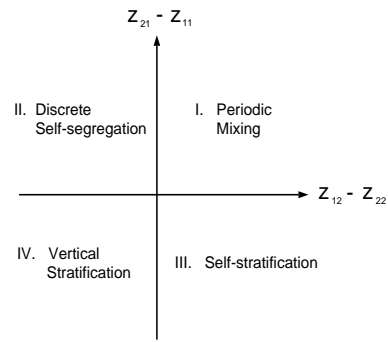,width=2in}}
\caption{Schematic diagram of \mbox{$z_{21}-z_{11}$}
versus \mbox{$z_{12}-z_{22}$} for \mbox{$N=2$}
and \mbox{$z_{11}<z_{22}$},
showing where each class of sandpile solution applies.
}
\label{f:param}
\end{figure}

A diagram of the parameter space showing the regions in which each
of these classes occurs is given in Fig.\ref{f:param}.
The borders between these regions correspond to
\mbox{$z_{21}=z_{11}$} or \mbox{$z_{12}=z_{22}$},
when domain stability is not well defined.
In these cases the sandpile behaviour is
either indeterminate between the two classes in question
or just reduces to random mixing.
For \mbox{$z_{11}=z_{22}$} periodic mixing and vertical
stratification are unaffected but there is no longer any
distinction between discrete self-segregation and self-stratification.
Instead, these two classes are replaced with a hybrid class
that exhibits self-segregation with a broad, layered boundary.

Each class is said to be {\em robust} if its existence is insensitive 
to the exact choice of dynamical rules.
It is possible to vary the volumes of each particle type added or
to introduce an open or closed right-hand boundary condition
without any significant alteration in the resultant sandpile.
Similarly, initially adding the particles over a range of sites
does not affect the sandpile to the right of the range,
and introducing annealed disorder to the 
$z_{\alpha\beta}$ just increases the noise.
More significant is the effect of
averaging the $z_{\alpha\beta}$ over adjacent pairs of sites,
corresponding perhaps to the nestling of the upper particle
in between the two lower ones, which destroys vertical
stratification and instead gives random mixing or discrete
self-segregation.
Crudely modelling inertia by allowing the moving particle to
stochastically drift a short distance further than normal
distorts vertical stratification and
replaces layering and periodic mixing with random mixing.
We conclude that the model is robust except when we
include displacements in the particle's horizontal motion.

For $N>2$ the parameter space for $z_{\alpha\beta}$
becomes too large to explore systematically.
The situation improves somewhat if only
domain stability is considered, but this still leaves
\mbox{$2^{N(N-1)}$} possible combinations, so
we have limited ourselves to a brief survey of all these cases for
$N=3$ and a representative sample for $N=4$.
The resultant sandpiles are essentially just
combinations of the four classes identified for $N=2$, the
only significant new feature being that periodic mixing can now
occur with periodicity \mbox{$\leq N$}.
Any particle separation and the allowed orders of periodic mixing
can be predicted in each case by extending the particle 
addition operator \mbox{$P_{\alpha\beta}(z_{i})$} to include
\mbox{$\alpha,\beta\in[1,N]$}.
As before, there is no generic way of constructing global solutions
but in many cases known solutions for $N-1$ particle systems can be used
instead.
This is possible when particle types~$\alpha$ and $\beta$
would by themselves periodically mix, when to good approximation
they can be replaced by a single particle type~$\alpha'$
which alternates between the two.
A domain is stable to $\alpha'$ only if it is stable to both
$\alpha$ and $\beta$, a domain of type~$\alpha'$ particles
is stable to another type only if both $\alpha$ and $\beta$ are, and
\mbox{$z_{\alpha'\alpha'}=\rm{min}(z_{\alpha\beta},z_{\beta\alpha})$}.
This reduced system usually exhibits the correct qualitative
structure of the original but significantly underestimates the
amount of noise.

\section{discussion}
\label{sec:disc}

The discrete model studied here is perhaps the simplest conceivable
model describing slowly driven sandpile formation with granular mixtures.
Nonetheless it exhibits a wide variety of non-trivial behaviour
in one-dimension, and we can only suppose that it will continue
to do so in higher dimensions.
The behaviour of sandpiles for binary mixtures falls
into one of four classes, two of which have known counterparts in the
rolling layer model~\cite{strat1,strat2}.
They are also much more susceptible to statistical fluctuations in
the order of particle types added.

We have recently initiated a series of experiments
in an attempt to reproduce these classes with real granular
materials~\cite{inprep}.
On first inspection, there may appear to be little hope that such
a simple model could describe real sandpile mixtures.
For instance, the
possibility of particles bouncing or dislodging surface
material has not been catered for.
However, our numerical
investigations into the robustness of the model leads
us to suppose that agreement might be possible in the limit of
overdamped particle motion.
It is also important to realise the limit of infinitely
slow driving as closely as possible, to minimise the probability
of a rolling layer forming, because even a thin
rolling layer would displace surface material along
the length of the slope. This would interfere with the formation of
both vertical stratification and periodic mixing.

In summary, sandpile formation by granular mixtures exhibits a greater
diversity of behaviour in the slowly driven limit than in the rolling
layer case, at least numerically.
If toppling were included into this model~\cite{topple},
more than one particle would be able to move simultaneously
in the form of an avalanche.
Similarly, allowing moving particles to dislodge surface
material in some manner might allow for something akin
to a rolling layer to form.
It would be interesting to see if the system diversity was
reduced in either of these two cases.

\section*{acknowledgement}
\label{sec:ack}

We would like to thank Peter Hobson for useful discussions.

\end{multicols}


\begin{references}

\bibitem [*] {em1} Electronic address:
David.Head@brunel.ac.uk

\bibitem [\dag] {em2} Electronic address:
G.J.Rodgers@brunel.ac.uk

\bibitem{granular} H.M.Jaeger and S.R.Nagel, Science {\bf 255}, 1523(1992).

\bibitem{brazilnut} A.Rosato, K.J.Strandburg, F.Prinz and R.H.Swendsen,
Phys.Rev.Lett. {\bf 58}, 1038(1987).

\bibitem{bn2} J.B.Knight, H.M.Jaeger and S.R.Nagel, Phys.Rev.Lett.
{\bf 70}, 3728(1993).

\bibitem{bn3} W.Cooke, S.Warr, J.M.Huntley and R.C.Ball,
Phys.Rev.E {\bf 53}, 2812(1996).

\bibitem{rotating} K.M.Hill and J.Kakalios, Phys.Rev.E {\bf 49},
R3610(1994).

\bibitem{rotating2} O.Zik, D.Levine, S.G.Lipson, S.Shtrikman
and J.Stavans, Phys.Rev.Lett. {\bf 73}, 644(1994).

\bibitem{strat1} H.A.Makse, S.Havlin, P.R.King and H.E.Stanley, submitted to 
Nature.

\bibitem{BTW} P.Bak, C.Tang and K.Wiesenfeld, Phys.Rev.A {\bf 38}, 364(1988).

\bibitem{Grinstein} G.Grinstein in {\em Scale Invariance, Interfaces and
Non-Equilibrium Dynamics}, Vol. 344 of {\em NATO Advanced Study Institute,
Series B:Physics}, edited by A.McKane {\em et.al.} (Plenum, New York, 1995).

\bibitem{exp1} H.M.Jaeger, C.Liu and S.R.Nagel, Phys.Rev.Lett.
{\bf 62}, 40(1989).

\bibitem{exp2} G.A.Held, D.H.Solina, II, D.T.Keane, W.J.Haag,
P.M.Horn and G.Grinstein, Phys.Rev.Lett. {\bf 65}, 1120(1990).

\bibitem{exp3} P.Evesque and J.Rajchenbach, Phys.Rev.Lett.
{\bf 62}, 44(1989).

\bibitem{ricepile} V.Frette, K.Christensen, A.Malthe-S\o renssen,
J.Feder, T.J\o ssang and P.Meakin, Nature {\bf 379}, 49(1996).

\bibitem{BCEP} J.-P.Bouchad, M.E.Cates, J.R.Prakash and S.F.Edwards,
Phys.Rev.Lett. {\bf 74}, 1982(1995).

\bibitem{strat2} H.A.Makse, P.Cizeau and H.E.Stanley, submitted to 
PRE or PRL.

\bibitem{footnote} This result can also be derived by a geometric
argument if it is assumed that both domains sweep out equal areas
in time,
corresponding to the equal rates of particle addition.

\bibitem{inprep} D.A.Head, P.R.Hobson and G.J.Rodgers, in preparation.

\bibitem{topple} L.P.Kadanoff, S.R.Nagel, L.Wu and S.Zhou,
Phys.Rev.A {\bf 39} 6524(1989).

\end{references}
\end{document}